\documentclass[aps,prl,twocolumn,amsmath,amssymb,superscriptaddress,nofootinbib]{revtex4-1}

\usepackage{aas_macros}
\usepackage{bm}
\usepackage{wasysym}

\begin{document}

\title{Lorentz Symmetry Violations from Matter-Gravity Couplings with Lunar Laser Ranging}

\author{A.~Bourgoin}
\affiliation{Dipartimento di Ingegneria Industriale, University of Bologna, via fontanelle 40, Forl\`i,  Italy}
\email{adrien.bourgoin@unibo.it}

\author{C.~Le~Poncin-Lafitte}
\affiliation{SYRTE, Observatoire de Paris, PSL Research University, CNRS, Sorbonne Universit\'es, UPMC Univ. Paris 06, LNE, 61 avenue de l'Observatoire, 75014 Paris, France}

\author{A.~Hees}
\affiliation{Department of Physics and Astronomy, University of California, Los Angeles, California 90095, USA}

\author{S.~Bouquillon}
\affiliation{SYRTE, Observatoire de Paris, PSL Research University, CNRS, Sorbonne Universit\'es, UPMC Univ. Paris 06, LNE, 61 avenue de l'Observatoire, 75014 Paris, France}

\author{G.~Francou}
\affiliation{SYRTE, Observatoire de Paris, PSL Research University, CNRS, Sorbonne Universit\'es, UPMC Univ. Paris 06, LNE, 61 avenue de l'Observatoire, 75014 Paris, France}

\author{M.-C.~Angonin}
\affiliation{SYRTE, Observatoire de Paris, PSL Research University, CNRS, Sorbonne Universit\'es, UPMC Univ. Paris 06, LNE, 61 avenue de l'Observatoire, 75014 Paris, France}

\date{\today}


\begin{abstract}
The standard-model extension (SME) is an effective field theory framework aiming at parameterizing any violation to the Lorentz symmetry (LS) in all sectors of physics. In this letter, we report the first direct experimental measurement of SME coefficients performed simultaneously within two sectors of the SME framework using lunar laser ranging observations. We consider the pure gravitational sector and the classical point-mass limit in the matter sector of the minimal SME. We report no deviation from general relativity and put new realistic stringent constraints on LS violations improving up to three orders of magnitude previous estimations.
\end{abstract}

\maketitle

\emph{Introduction.---}At the classical level, general relativity (GR) is known to describe accurately the gravitational phenomenons over a wide range of distance scales \cite{2014LRR....17....4W}. At the quantum level, the standard-model of particle physics is also a great success of modern physics. It incorporates the laws of special relativity into a quantum field theory offering an accurate description of matter and non-gravitational forces. These two pillars of modern physics provide a deep understanding in the description of Nature based on a unique symmetry of space-time known as the Lorentz symmetry (LS).

However, merging GR with the standard-model of particles in a single unified theory remains a challenging task. Actually, GR is a classical field theory describing the gravitational interaction as the classical consequence of space-time curvature induced by its matter-energy content. On the other hand, quantum field theory describes electromagnetic, weak, and strong interactions with the quantum exchange of subatomic particles. In an attempt to construct a quantum gravity theory many scenarios have been proposed. However, none of them have yet resulted in a completely satisfactory theory able to make testable predictions. For instance, the experimental effects are expected to become relevant at the Planck scale ($m_P=10^{19}$ GeV), where they may manifest as tiny LS violations \cite{1989PhRvD..39..683K}. Such a high energy level is nowadays unreachable; however at lower energy, high precision experiments should be able to detect these LS violations if they exist \cite{2005LRR.....8....5M,2014RPPh...77f2901T}.

In this context, an effective field theory, the standard-model extension (SME), was constructed to consider and classify LS violations in all sectors of physics \cite{1995PhRvD..51.3923K,1997PhRvD..55.6760C,*1998PhRvD..58k6002C,2004PhRvD..69j5009K}. SME contains both the standard-model and GR Lagrangians and include all possible Lorentz-violating terms in all sectors of physics. Considering the wide range of applicability of this formalism, there exist a lot of parameters to be determined by many different types of experiments~\cite{2011RvMP...83...11K}. In the following, we will focus on two aspects of the SME, namely the pure gravitational sector \cite{2006PhRvD..74d5001B} and the classical point-mass limit in the matter sector \cite{2011PhRvD..83a6013K} of the minimal SME.

Following from \cite{2006PhRvD..74d5001B,2011PhRvD..83a6013K}, an hypothetical breaking of the LS in the gravitational and matter sectors naturally leads to an expansion at the level of the action which is written $S_{\text{tot}}=S_g+S_m+S'$. The first term $S_g$ is the action of the gravitational field \cite{2006PhRvD..74d5001B}, containing a Lorentz invariant part (the Einstein-Hilbert action of GR) and an additional Lorentz-violating part which includes new LS violating fields contracted to gravitational field operators. The second term $S_m$ is the matter field of point-mass particles which is written at leading order as \cite{2011PhRvD..83a6013K}
\begin{equation}
  S_m=-mc\!\int\! \mathrm d\lambda\Big(\sqrt{(g_{\mu\nu}\!+\!c_{\mu\nu})u^{\mu}u^{\nu}}\!+\!\frac{1}{m}(a_{\text{eff}})_{\mu}u^{\mu}\Big)\text{,}
  \label{eq:Sm}
\end{equation}
with $\lambda$ an affine parameter, $u^{\mu}=\mathrm dx^{\mu}/\mathrm{d}\lambda$ the four-velocity of the particle, $m$ its mass, $c$ the speed of light in a vacuum, and $(a_{\text{eff}})_{\mu}$ and $c_{\mu\nu}$ are the Lorentz-violating fields. The last term $S'$ contains the dynamics associated with the Lorentz-violating fields.

Experimental evidences imply the Lorentz-violating fields to be small quantities \cite{2006PhRvD..74d5001B,2011PhRvD..83a6013K}. This justifies to consider the linearized gravity limit where the observables only depend on the vacuum expectations value of the Lorentz-violating fields (denoted $\bar c_{\mu\nu}$ and $(\bar a_{\text{eff}})_{\mu}$ for the gravity-matter couplings, and $\bar s^{\mu\nu}$ for the pure gravitational sector). All the coefficients $\bar s^{\mu\nu}$, $\bar c_{\mu\nu}$, and $(\bar a_{\text{eff}})_{\mu}$ control the Lorentz-violating effects at the level of the field equations. However, $\bar c_{\mu\nu}$ and $(\bar a_{\text{eff}})_{\mu}$ have also an important property since they are species dependent \cite{2011PhRvD..83a6013K}. Such a dependence in the action of point-mass particles [see Eq. \eqref{eq:Sm}] induces violations of the three facets of the Einstein equivalence principle (EEP) and leads to deviation of the geodesic motion depending at first order on the background values of the Lorentz-violating fields $\bar c_{\mu\nu}$ and $(\bar a_{\text{eff}})_{\mu}$.

In the past few years, the $\bar s^{\mu\nu}$ coefficients alone have been extensively investigated in postfit analyzes (based on theoretical grounds) with various techniques \cite{2008PhRvL.100c1101M,*2009PhRvD..80a6002C,2013PhRvD..88j2001B,2014PhRvL.112k1103S,*2014PhRvD..90l2009S,2012CQGra..29q5007I,2015PhRvD..92f4049H,2015PhLB..749..551K,2016PhLB..757..510K,2007PhRvL..99x1103B,sym8110111,2017arXiv170702318S}. However, the most stringent constraints were obtained in global data processing (direct experimental measurement) from very long baseline interferometry (VLBI) \cite{2016PhRvD..94l5030L} and lunar laser ranging (LLR) \cite{PhysRevLett.117.241301}. Concerning the two other sets of coefficients related to the matter-gravity couplings, $\bar c_{\mu\nu}$ is the most extensively considered \cite{PhysRevLett.106.151102,PhysRevLett.111.151102,PhysRevLett.96.060801,PhysRevLett.111.050401,2017PhRvD..95g5026P}. On the opposite, the $(\bar a_{\text{eff}})_{\mu}$ are sparsely constrained from postfit analyzes with atom interferometry, planetary and lunar ephemerides \cite{2015PhRvD..92f4049H}, nuclear bending energy \cite{PhysRevLett.111.151102}, and superconducting gravimeters \cite{2016arXiv161208495F}.

As discussed in \cite{2016PhRvD..94l5030L,PhysRevLett.117.241301,2016Univ....2...30H}, constraints derived in postfit analyzes are not fully satisfactory since all the correlations between the SME coefficients and the other global parameters (masses, initial conditions and so on) are neglected which leads to overoptimistic errors in the estimated parameters. Realistic constraints can only be determined in a global data processing where all the correlations are considered. Furthermore, postfit analyzes rely on analytical signatures derived using simplifying assumptions leading to a loss of precision in the determination of the sensitivities to SME coefficients \cite{PhysRevLett.117.241301}. 

In this letter, we derive realistic estimates for both SME coefficients $\bar s^{\mu\nu}$ and $(\bar a_{\text{eff}})_{\mu}$ simultaneously from a global LLR data processing, neglecting the already very well constrained coefficients $\bar c_{\mu\nu}$ (see Refs. \cite{PhysRevLett.106.151102,PhysRevLett.111.151102,PhysRevLett.96.060801,PhysRevLett.111.050401}).

\emph{LLR experiment.---}The LLR is an astrometric experiment devoted to the accurate timing of the round trip of short laser pulses between a LLR station on Earth and a retroreflector corner cube at the Moon's surface. In term of distance, the current precision reaches the subcentimetric level \cite{1998A&AS..130..235S,2013RPPh...76g6901M}. During the last 48 years, the data were acquired by five LLR stations namely McDonald Observatory in Texas, Observatoire de la C\^ote d'Azur in France, Haleakala Observatory in Hawaii, Apache Point Observatory in New Mexico, and Matera in Italy. The Apache Point Observatory is dedicated to millimetric measurements since 2006 \cite{2013RPPh...76g6901M} and realizes the most accurate observations in the green wavelength to date. Since March 2015, the infrared (IR) wavelength is preferred at the Calern site in France due to a better atmospheric transmission \cite{2017LPI....48.2329V}.

At the lunar surface, the laser pulses are currently reflected by five retroreflectors; three were installed by the Apollo missions XI, XIV, and XV, and the other two were put on the Soviet rovers Lunokhod 1 and 2.

The data is distributed as normal points which contain light travel time of photons averaged over several minutes in order to achieve a higher signal to noise ratio measurement. The variations of the round trip travel time contain a lot of information about physical properties of the Earth-Moon system. These observations can be used, among others, to probe fundamental properties of gravitation like e.g. the LS. 

\emph{Numerical modeling.---}In this Letter, we simulate numerically, at the subcentimetric level, each LLR normal point using a modeling developed within the SME framework. The theoretical expression of the LLR observable is thoroughly presented in \cite{PhysRevLett.117.241301} [see Eqs. (2) and (3)]. It depends on many different contributions, such as the geometrical distance between the Earth and the Moon (that depends directly on the equations of motion), the gravitational time delay, the atmospheric delay, etc\dots To account for LS breaking, we have built a new numerical lunar ephemeris named ``\'Eph\'em\'eride Lunaire Parisienne Num\'erique'' (ELPN) which computes the orbital and rotational motions of the Moon in the SME framework. This ephemeris takes into account all the physical effects which produce a signal larger than the millimeter over the Earth-Moon distance. A first version of ELPN including only the pure gravitational sector of the minimal SME has already been presented in \cite{PhysRevLett.117.241301}. In this work, we extend the theoretical framework of ELPN to include a breaking of LS in the gravity-matter sector from the SME framework. A breaking of LS will affect the LLR observable at two different levels: (i) it will modify the orbital motion and (ii) it will modify the gravitational time delay.

The SME contribution to the equations of motion is given by~\cite{2006PhRvD..74d5001B,2011PhRvD..83a6013K,2015PhRvD..92f4049H}
\begin{widetext}
  \begin{align}
  a^J=\frac{G_{\!N}M}{r^3}&\Bigg\{\bar s^{JK}_tr^K-\frac{3}{2}\bar s^{KL}_t\hat r^K\hat r^Lr^J-\bar s^{TJ}\hat V^Kr^K-\bar s^{TK}\hat V^Jr^K+3\bar s^{TL}\hat V^K\hat r^K\hat r^Lr^J+3\bigg[\bar s^{TK}-\cfrac{2}{3}\sum_w\cfrac{n_3^w}{M}\alpha(\bar a_{\text{eff}}^w)^K\bigg]\hat V^Kr^J\Bigg.\nonumber\\
  &\Bigg.+2\frac{\delta m}{M}\bigg[\bar s^{TK}+\sum_w\cfrac{n_2^w}{\delta m}\alpha(\bar a_{\text{eff}}^w)^K\bigg]\hat v^Kr^J-2\frac{\delta{}m}{M}\bigg[\bar s^{TJ}+\sum_w\cfrac{n_2^w}{\delta m}\alpha(\bar a_{\text{eff}}^w)^J\bigg]\hat v^Kr^K\Bigg\}\text{,}\label{eq:LLR_eom}
  \end{align}
\end{widetext}
where $M=m_{\leftmoon}+m_\oplus$, $\delta m=m_\oplus-m_{\leftmoon}$, $n_2^w = N^w_{\leftmoon}-N^w_\oplus$, $n_3^w =M(N^w_{\leftmoon}/m_{\leftmoon}+N^w_\oplus/m_\oplus)$, $\hat r^J=r^J/r$, $\hat v^J=v^J/c$, and $\hat V^J=V^J/c$.

In this expression, $r^J$ is the position of the Moon with respect to the Earth, $v^J$ is the relative velocity of the Moon with respect to the Earth, $V^K$ is the heliocentric velocity of the Earth-Moon Barycenter, and $N^w$ is the number of particles of species $w$ (electrons, protons and neutrons). We used the three-dimensional traceless tensor $\bar s^{JK}_t=\bar s^{JK}-\frac{1}{3}\bar s^{TT}\delta^{JK}$ and a rescaled observable Newtonian constant defined as $G_N = G\left(1+\frac{5}{3}\bar s^{TT}\right)$ \cite{2013PhRvD..88j2001B}. One can notice that this perturbing acceleration is  dependent on the composition of the bodies meaning that the EEP is violated. This equation has been implemented with all the associated partial derivatives in ELPN. 

Concerning the gravitational time delay denoted by $\Delta\tau_g$, we have also considered the gravity-matter couplings from the SME \cite{2011PhRvD..83a6013K}. This contribution, once added to the pure gravitational sector of the minimal SME \cite{2006PhRvD..74d5001B,2009PhRvD..80d4004B}, leads to the following expression for the one-way gravitational time delay
\begin{widetext}
  \begin{align}
    \Delta\tau_g(\bm x_e, \bm x_r)=-\sum_{b=\odot,\oplus}&\Bigg\{\frac{2 G_{\!N}m_b}{c^3}\bigg[\bar s^{TJ}-\sum_w\frac{N_b^w}{m_b}\alpha(\bar a^w_{\text{eff}})^J\bigg]\hat{x}^J_{er}\ln\Bigg[\frac{r_{be}+r_{br}+r_{er}}{r_{be}+r_{br}-r_{er}}\Bigg]\nonumber\\
    &\Bigg.-\frac{{G}_{\!N}m_b}{c^3}\Bigg(\bar s_t^{JK}\hat{p}_b^J\hat{p}_b^K+\bigg[\bar s^{TJ}-\sum_w\frac{N_b^w}{m_b}\alpha(\bar a^w_{\text{eff}})^J\bigg]\hat{x}^J_{er}\Bigg)\Big(\hat{x}_{br}^K\hat{x}^K_{er}-\hat{x}_{be}^K\hat{x}^K_{er}\Big)\nonumber\\
    &\Bigg.+\frac{{G}_{\!N}m_b}{c^3}\Bigg(\bar s_t^{JK}\hat{x}^J_{er}\hat{p}_b^K-\bigg[\bar s^{TJ}-\sum_w\frac{N_b^w}{m_b}\alpha(\bar a^w_{\text{eff}})^J\bigg]\hat{p}_b^J\Bigg)\frac{p_b\big(r_{be}-r_{br}\big)}{r_{be}r_{br}}\Bigg\}\text{,}\label{eq:dt}
  \end{align}
\end{widetext}
where $\bm x_{e/r}$ are the positions of the emitter and receiver (the LLR station on Earth and one of the lunar reflector depending on the way of the signal), $\bm x_{er}=\bm x_r - \bm x_e$, $\bm x_{br}=\bm x_r-\bm x_b$, $\bm x_{be}=\bm x_e-\bm x_b$ with $\bm x_b$ the position of the source that generates gravitation (here the Earth or the Sun), the hat refers to vectors that are normalized so that they are unit vectors and $p_b^J$ is the impact parameter vector defined by $\bm p_b=\hat {\bm x}_{er} \times \left(\bm x_{br}\times\hat {\bm x}_{er}  \right) $~\cite{2009PhRvD..80d4004B}.

We have used a  model for the composition of the Sun characterized by $N_{\odot}^e/m_{\odot}=N_{\odot}^p/m_{\odot}\approx0.9(\text{GeV}/c^2)^{-1}$ and $N_{\odot}^n/m_{\odot}\approx0.1(\text{GeV}/c^2)^{-1}$ \cite{2011PhRvD..83a6013K}. For the composition of the Earth, we have considered that $N_{\oplus}^e/m_{\oplus}=N_{\oplus}^p/m_{\oplus}\approx N_{\oplus}^n/m_{\oplus}\approx0.5(\text{GeV}/c^2)^{-1}$ (idem for the Moon), as in \cite{2015PhRvD..92f4049H,PhysRevLett.70.2220}. It is worth mentioning that considering LS violations induced by neutral macroscopic bodies (with equal number of electrons and protons) makes the data being only sensitive to a combination involving electrons and protons like $\alpha(\bar a^{e+p}_{\text{eff}})_{\mu}=\alpha(\bar a^e_{\text{eff}})_{\mu}+\alpha(\bar a^p_{\text{eff}})_{\mu}$.

We do not mention terms in $\bar s^{TT}$ and $\alpha(\bar a_{\text{eff}}^w)^T$ in Eq. \eqref{eq:dt}. They are absorbed in $G_N$ in the orbital part [see Eq. \eqref{eq:LLR_eom}], so they are only supposed to show up in the gravitational light time delay expression. Unfortunately, considering the current accuracy of LLR data, a simple computation using Eq. \eqref{eq:dt} reveals that only an upper limit of $10^{-2}$ -- $10^{-3}$ can be reached for a single combination involving $\bar s^{TT}$ and $\alpha(\bar a_{\text{eff}}^w)^T$. Such limits are not competitive with other determinations from VLBI \cite{2016PhRvD..94l5030L}, nuclear bending energy \cite{PhysRevLett.111.151102}, or atom interferometry \cite{PhysRevLett.106.151102}.

The numerical integration of the equations of motion gives the time evolution of the Earth-Moon distance, the lunar librations, and all the associated partial derivatives. Then, these quantities are transformed into a theoretical round-trip light time following the IERS standard 2010 \cite{2010ITN....36....1P}.  The residuals are deduced by comparing these theoretical estimations with the measurements and are finally minimized with a standard iterative least-square fit.

\emph{Solution in GR framework.---}The procedure followed in this analysis is similar to the one described in \cite{PhysRevLett.117.241301}. In a first step, we have built a reference solution in pure GR by imposing the nullity of the Lorentz-violating coefficients. The initial physical parameters and initial conditions are taken from DE430 \cite{2014IPNPR.196C...1F}. Then, the independent solution ELPN is built with an iterative process consisting of adjusting 59 parameters to 24022 normal points spanning 48 years of LLR observations from August 1969 to December 2016 (these parameters include e.g. the position of the LLR stations and retroreflectors at J2000, the orbital and rotational lunar initials conditions at J2000, the masses of the Moon and the Earth-Moon barycenter, the Love's numbers and the time delays of degree 2 for solid body tides of both the Earth and the Moon, the total moment of inertia of the Moon, and the damping term between the mantle and the fluid core of the Moon). Among these normal points we have considered 1337 observations in IR wavelength obtained by the Grasse station in France \cite{2017arXiv170406443C}. The dispersion of the residuals obtained with ELPN in pure GR at the end of the iterative process are shown in Tab. \ref{tab:res}. They reveal that our numerical solution is perfectly accurate, especially for the most recent observations (Apache-point and Grasse [MeO, IR]) for which the dispersion is at the 2cm level.

\begin{table}[b]
  \begin{tabular}{l c r r r}
    \toprule
    station/instrument  & Period   & \multicolumn{1}{c}{$N$} &  \multicolumn{1}{c}{$\mu$[$cm$]} & \multicolumn{1}{c}{$\sigma$[$cm$]} \\
    \hline
    Haleakala               & 1984-1990 & 770   & -0.6  & 10.4 \\
    Matera                  & 2003-2015 & 118   & -1.0  & 8.9  \\
    McDonald ($2.7m$)       & 1969-1985 & 3604  & 9.1   & 34.9 \\
    McDonald (MLRS1)        & 1983-1988 & 631   & 5.2   & 37.9 \\
    McDonald (MLRS2)        & 1988-2015 & 3670  & 0.4   & 9.0  \\
    Grasse (Rubis)          & 1984-1986 & 1188  & 4.7   & 18.2 \\
    Grasse (Yag)            & 1987-2005 & 8324  & -0.5  & 4.7  \\
    Grasse (MeO)            & 2009-2016 & 1732  & 0.0   & 2.4  \\
    Grasse (IR)             & 2015-2016 & 1337  & -1.5  & 2.2  \\
    Apache-point            & 2006-2010 & 941   & 0.0   & 2.6  \\
    Apache-point            & 2010-2012 & 513   & 0.0   & 3.3  \\
    Apache-point            & 2012-2013 & 360   & 0.0   & 3.2  \\
    Apache-point            & 2013-2016 & 834   & 0.0   & 2.1  \\
    \botrule
  \end{tabular}
  \caption{Residuals of ELPN in pure GR per LLR stations and instruments. $\mu$ is the mean of the dispersion and $\sigma$ is the dispersion around the mean. For each station/instrument, $N$ is the number of available observations.}
  \label{tab:res}
\end{table}

This new GR solution constitutes the starting point of our analysis of LS violation. Adopting the same procedures as in the literature \cite{2011RvMP...83...11K,2016PhRvL.117g1102S,2016arXiv161208495F}, we extract LS violation sensitivities using two methods. The most rigorous one is called ``coefficient separation'' and the other one is the ``maximum reach''.

\emph{Coefficient separation.---}In this procedure, all the Lorentz-violating fields are treated as nonzero simultaneously. A global fit with the other 59 physical parameters shows that some correlations between SME coefficients are very high (larger than 95\%), meaning that the data is sensitive to linear combinations of the SME coefficients. An iterative analysis of partial derivatives allows us to determine the most sensitive and independent linear combinations attainable with the LLR experiment
\begin{subequations}\label{eq:comb}
  \begin{align}
    \bar s^1&=\bar s^{XY}\text{,}\allowdisplaybreaks\label{eq:combi1}\\
    \bar s^2&=\bar s^{XZ}\text{,}\allowdisplaybreaks\label{eq:combi2}\\
    \bar s^3&=\bar s^{XX}-\bar s^{YY}\text{,}\label{eq:combi3}\\
    \bar s^4&=0.35\bar s^{XX}+0.35\bar s^{YY}-0.70\bar s^{ZZ}-0.94\bar s^{YZ}\text{,}\allowdisplaybreaks\label{eq:combi4}\\
    \bar s^5&=-0.62\bar s^{TX}+0.78\alpha(\bar a_{\text{eff}}^{e+p})^X+0.79\alpha(\bar a_{\text{eff}}^{n})^X\text{,}\label{eq:combi5}\\
    \bar s^6&=0.93\bar s^{TY}+0.34\bar s^{TZ}-0.10\alpha(\bar a_{\text{eff}}^{e+p})^Y-0.10\alpha(\bar a_{\text{eff}}^{n})^Y\nonumber\\
    &-0.044\alpha(\bar a_{\text{eff}}^{e+p})^Z-0.044\alpha(\bar a_{\text{eff}}^{n})^Z\text{.}\label{eq:combi6}
  \end{align}
\end{subequations}
At the end, two fundamental SME coefficients ($\bar s^{XY}$ and $\bar s^{XZ}$) and four linear combinations ($\bar s^{3-6}$) can be estimated without high correlations (the largest one remains below 30\%). A fit including the linear combinations with the 59 physical parameters provide estimations of Eqs.~\eqref{eq:comb} with their statistical uncertainties.

It is known that the least-square fit returns only statistical uncertainties (labeled $\sigma_{\text{stat}}$) which may be overoptimistic since no systematic is considered. However, in LLR observations all the data acquired by one instrument is not independent, which results in systematic effects in the estimations of parameters. Therefore, it is essential to quantify such neglected systematics in order to provide realistic uncertainties on the estimated parameters.

In order to assess these systematics we used a Jackknife resampling method (see e.g. \cite{1993stp..book.....L} and \cite{PhysRevLett.117.241301} for a similar use of this method to LLR data). We built 18 subsets of data: 13 by station/instrument (as depicted in Tab.~\ref{tab:res}) and 5 by retroreflectors. Each subset is constructed by removing all the observations acquired by one station/instrument or reflected by one retroreflector. The basic idea is to consider that each subset provides an independent estimation of the SME coefficients which is used to infer a systematic uncertainty (for more details see \cite{PhysRevLett.117.241301}). We have applied this procedure to (i) subsets by stations/instruments (the obtained systematics variance is labeled $\sigma_s^2$) and to (ii) subsets by retroreflectors (the obtained systematics variance is labeled $\sigma_r^2$).

Finally, the total variance estimate is the sum of statistical and the two systematics uncertainties $\sigma^2=\sigma_{\text{stat}}^2+\sigma_{\text{syst}}^2$ with $\sigma_{\text{syst}}^2=\sigma_s^2+\sigma_r^2$. The final estimations with the associated realistic errors on SME coefficients are given in Tab. \ref{tab:est}. No deviation from GR  is reported.

\begin{table}[b]
  \begin{tabular}{c l c}
    \toprule
    \multicolumn{1}{c}{SME} & \multicolumn{1}{c}{Constraints} & \multicolumn{1}{c}{$|{}^{\sigma_{\text{syst}}}/{}_{\sigma_{\text{stat}}}|$} \\
    \hline
    $\bar s^1$ & $(-0.5\pm3.6)\times 10^{-12}$ & 3.7 \\
    $\bar s^2$ & $(+2.1\pm3.0)\times 10^{-12}$ & 8.8 \\
    $\bar s^3$ & $(+0.2\pm1.1)\times 10^{-11}$ & 2.8 \\
    $\bar s^4$ & $(+3.0\pm3.1)\times 10^{-12}$ & 4.5 \\
    $\bar s^5$ & $(-1.4\pm1.7)\times 10^{-8}$ & 4.8 \\
    $\bar s^6$ & $(-6.6\pm9.4)\times 10^{-9}$ & 4.1 \\
    \botrule
  \end{tabular}
  \caption{Realistic constraints on SME linear combinations [see Eqs. \eqref{eq:comb}] from a global LLR data analysis in ``coefficient separation'' approach. The quoted uncertainties correspond to $1\sigma$ realistic uncertainties.}
  \label{tab:est}
\end{table}

Our results improve up to a factor 2 previous estimations in the pure gravitational sector \cite{2016Univ....2...30H} on $\bar s^{XY}$, $\bar s^{XZ}$, and $\bar s^3$. However, the estimation on the combination $\bar s^4$ is improved by more than one order of magnitude compared to \cite{PhysRevLett.117.241301}. This improvement is mainly due to the consideration of 3300 additional subcentimetric data distributed between December 2013 and December 2016.

The main novelty from this work is related to the last two linear combinations ($\bar s^5$ and $\bar s^6$) which regroup the gravity-matter couplings coefficients $(\bar a_{\text{eff}})^J$ with the boost coefficients $\bar s^{TJ}$. In previous studies based on theoretical grounds, the $(\bar a_{\text{eff}})^J$ coefficients were shown to appear in four linear combinations instead of two (see Eqs. (20) of \cite{2015PhRvD..92f4049H}), highlighting the limitations of postfit methods. Such a difference is explained since we have numerically integrated effects of LS violations, considering in this way, short and long-term signatures. The main result rely on the bounds over $\bar s^5$ and $\bar s^6$ (cf. Tab.~\ref{tab:est}) which are at the level of one part in $10^8$ representing an improvement of almost two orders of magnitude compared to previous best determination~\cite{2015PhRvD..92f4049H} (considering results from the LLR section).

\emph{Maximum reach.---}This procedure is less general than the previous one and is based on the assumption that no set of the SME coefficients could be generated in the underlying theory in such a way that they lead to an exact cancellation in observable effects \cite{2016arXiv161208495F}. In this approach, each Lorentz-violating coefficient is estimated separately assuming all the others vanish. We have performed successively a global fit of the 59 physical parameters with each SME coefficient one by one. Then, we have deduced realistic errors performing the same Jackknife resampling method as the one discussed in the previous section. Final estimations are presented in Tab. \ref{tab:est2}. No deviation from GR  is reported.

Our results improve (except $\bar s^{TX}$) previous best estimations in the pure gravitational sector \cite{2016Univ....2...30H} up to a factor 2. However, in the matter sector, the improvement is global compared to the best current postfit determinations \cite{2016arXiv161208495F,2015PhRvD..92f4049H}. In particular, we improve the constraints on the matter sector coefficients by two [for $\alpha(\bar a_{\text{eff}}^{n})^X$] to three orders of magnitude [for $\alpha(\bar a_{\text{eff}}^{e+p})^Z$]. In addition, let us notice that our constraints take into account correlations with the other global physical parameters where the estimations deduced with postfit methods does not.

We insist on the fact that results obtained with this approach are less general than the ones obtained with the ``coefficient separation'' method. 

\begin{table}
  \begin{tabular}{c l c}
    \toprule
    \multicolumn{1}{c}{SME} & \multicolumn{1}{c}{Constraints} & \multicolumn{1}{c}{$|{}^{\sigma_{\text{syst}}}/{}_{\sigma_{\text{stat}}}|$} \\
    \hline
    $\bar s^{XX}-\bar s^{YY}$             & $(-1.1\pm7.1)\times 10^{-12}$ & 4.5 \\
    $\bar s^{XX}+\bar s^{YY}-2\bar s^{ZZ}$ & $(+2.0\pm2.8)\times 10^{-11}$ & 8.9 \\
    $\bar s^{XY}$                        & $(-1.9\pm3.4)\times 10^{-12}$ & 8.0 \\
    $\bar s^{XZ}$                        & $(+3.2\pm3.7)\times 10^{-12}$ & 12.3 \\
    $\bar s^{YZ}$                        & $(-4.1\pm4.6)\times 10^{-12}$ & 7.4 \\
    $\bar s^{TX}$                        & $(+1.5\pm1.6)\times 10^{-8}$ & 6.4 \\
    $\bar s^{TY}$                        & $(+0.3\pm5.2)\times 10^{-9}$ & 4.7 \\
    $\bar s^{TZ}$                        & $(-0.5\pm7.7)\times 10^{-9}$ & 4.1 \\
    $\alpha(\bar a_{\text{eff}}^{e+p})^X$   & $(-2.5\pm2.9)\times 10^{-8}$ & 6.9 \\
    $\alpha(\bar a_{\text{eff}}^{e+p})^Y$   & $(+0.4\pm1.5)\times 10^{-8}$ & 3.1 \\
    $\alpha(\bar a_{\text{eff}}^{e+p})^Z$   & $(+2.4\pm3.6)\times 10^{-8}$ & 3.4 \\
    $\alpha(\bar a_{\text{eff}}^{n})^X$    & $(-2.5\pm2.9)\times 10^{-8}$ & 6.8 \\
    $\alpha(\bar a_{\text{eff}}^{n})^Y$    & $(+0.4\pm1.5)\times 10^{-8}$ & 3.1 \\
    $\alpha(\bar a_{\text{eff}}^{n})^Z$    & $(+2.4\pm3.6)\times 10^{-8}$ & 3.4 \\  
    \botrule
  \end{tabular}
  \caption{Realistic constraints on SME coefficients from a global LLR data analysis in ``maximum reach'' approach. The quoted uncertainties correspond to $1\sigma$ realistic uncertainties.}
  \label{tab:est2}
\end{table}

\emph{Conclusion.---}In this letter, we presented a simultaneous determination of the pure gravitational and matter-gravity coupling SME coefficients using LLR observations. Our results improve current constraints by up to three orders of magnitude. A key point is addressed when no assumption is assumed on the exact cancellation of SME coefficients. In that case, LLR data is only sensitive (through $\bar s^5$ and $\bar s^6$) to a combination of the two SME sectors and does not allow to disentangle them, meaning that LLR does not provide a pure test of the EEP (already pointed out by \cite{1988PhRvD..37.1070N}). To disentangle this ambiguity, our results in Tab.~\ref{tab:est} have to be combined with other measurements related to EEP (and not based on theoretical grounds but rather on direct measurement from experiment in order to maintain the robustness of the current analysis) like e.g. test of the universality of free fall with MicroSCOPE \cite{2001CQGra..18.2487T,2017arXiv170511015P} or tests of the gravitational redshift like e.g. with the Galileo V and VI GNSS satellites \cite{2015CQGra..32w2003D}.
  
\emph{Acknowledgments.---}Authors thank the LLR staff observers and the ILRS for the data collection and normal points distribution. A.B. thanks SYRTE in Paris Observatory for financial support, C.L.P.L. is grateful for the CNRS/GRAM and “Axe Gphys” of Paris Observatory Scientific Council.  
  
\bibliographystyle{apsrev4-1}
\bibliography{SME_GMC_LLR}

\end{document}